\documentclass[12pt]{article}
\usepackage[dvips]{graphicx}

\begin{document}
\title{Nyquist Method for Wigner-Poisson Quantum Plasmas}
\author{
F.~Haas\footnote{ferhaas@lncc.br \hskip 10pt
$^\dag$giovanni.manfredi@lpmi.uhp-nancy.fr \hskip 10pt
$^\ddag$goedert@exatas.unisinos.br}\\
Laborat\'orio Nacional de Computa\c{c}\~ao Cient\'{\i}fica - LNCC \\
Av. Get\'ulio Vargas, 333\\
25651-07 Petr\'opolis, RJ, Brazil\\
G.  Manfredi$^\dag$ \\
Laboratoire de Physique  des Milieux Ionis\'es, Universit\'e Henri Poincar\'e,\\
BP 239, 54506 Vandoeuvre-les-Nancy, France\\
J.~Goedert$^\ddag$\\
Centro de Ci\^encias Exatas e Tecnol\'ogicas - UNISINOS\\
Av. Unisinos, 950\\
93022-000 S\~ao Leopoldo, RS, Brazil}

\maketitle

\begin{abstract}
By means of the Nyquist method, we investigate the linear stability
of electrostatic waves in homogeneous equilibria of
quantum plasmas described by the Wigner-Poisson system. We show
that, unlike the classical Vlasov-Poisson system, the
Wigner-Poisson  case does not necessarily possess a Penrose
functional determining its linear stability properties.  The
Nyquist method is then applied to a two-stream distribution, for which we obtain
an exact, necessary and sufficient condition for linear
stability, as well as to a bump-in-tail equilibrium.
\end{abstract}


\section{Introduction}
The topic of quantum plasmas has recently attracted considerable
attention \cite{Kluksdahl}-\cite{Ancona}. A central reason for
this accrued  interest derives from the importance of quantum
effects in the performance of today's micro-electronic devices,
for which classical transport models are not always adequate in
view of the increasing  miniaturization level that is now entering
the submicron domain.  Hence, it is desirable to achieve a good
understanding of the basic properties of quantum transport models.
The Wigner-Poisson system \cite{Wigner}-\cite{Tatarski} is a
quantum transport model that has proven to be suitable in the
treatment of quantum devices like the resonant tunneling diode
\cite{Kluksdahl}. Moreover, it has been referred \cite{Markowich}
to as perhaps the only {\it kinetic} quantum transport model
amenable to detailed numerical simulation. In the present work, we
address the question of the stability of small-amplitude waves,
described by the Wigner-Poisson system.

A convenient tool to investigate the linear stability of systems
having a dispersion relation is provided by the Nyquist method
\cite{Krall},\cite{Penrose}.  Let us briefly review the basis of
this approach. Let $D(\omega,k) = 0$ be the dispersion relation,
where $\omega$ and $k$ are the frequency and wave-number for
small-amplitude oscillations. In most practical cases, it is
impossible to solve exactly the dispersion relation for $\omega$
as a function of $k$, some kind of approximation being necessary.
Hence, the imaginary part of the frequency, which determines the
stability properties of the system, can be obtained only in an
approximate way. However, exact results can be found by splitting
$D$ in its real and imaginary parts,  $D(\omega,k) =
D_{r}(\omega,k) + iD_{i}(\omega,k)$. Then, for fixed $k$ and real
$\omega$, by varying $\omega$ from minus to plus infinity we can
draw a diagram in the $D_{r} \times D_{i}$ plane. The resulting
curve, known as the Nyquist diagram, determines the number of
unstable modes of the system, which
equals the number of times the origin is encircled by the diagram
\cite{Krall}.
For example, using the Nyquist method, one can show that
equilibrium distributions that
are monotonically decreasing functions of the energy are stable against small
perturbations.  Moreover, for symmetric equilibria with at most
two maxima, the sign of the so-called Penrose functional
\cite{Krall},\cite{Penrose} determines the linear stability of the
classical Vlasov-Poisson system.

In view of the utility of Nyquist's method for classical plasmas,
it seems desirable to investigate whether it can be applied to the quantum
case too. This approach is
justified, since the linear stability of waves in the
Wigner-Poisson system is described by a dispersion relation, and
is therefore amenable to Nyquist's treatment.  However,
we cannot a priori expect to obtain a result as general as in the classical
case.  Indeed, as we shall see, the question of stability is
subtler in the quantum framework, a typical example being provided by
the two-stream instability \cite{Haas}. For simplicity, in the present work we
shall only consider homogeneous equilibria for one-dimensional
electrostatic plasmas consisting of mobile electrons. An immobile
ionic background guarantees overall charge neutrality.

The paper is organized as follows. In Section II, we develop the
fundamentals of the Nyquist method as applied to quantum plasmas
described by the Wigner-Poisson system. The stability properties
of quantum plasmas are determined by the specific form of the quantum
dispersion relation \cite{Drummond},\cite{Klimontovich}.  We show
that there are a rich variety of possible behaviors in quantum systems,
which are not present in classical Vlasov-Poisson plasmas. In
particular, in Section III, we prove that a
quantum analogue of the Penrose functional cannot exist.  To show this,
we consider symmetric equilibria with at most two maxima.
Nevertheless, the Nyquist method can still be used for Wigner-Poisson
plasmas. This is explicitly shown in Section IV, where we study a
two-stream equilibrium, described by a bi-Lorentzian distribution
function, which is amenable to exact calculations. We
find an exact criterion for stability, which reduces to the classical
criterion when quantum effects becomes negligible.  However,
large quantum effects can destroy the instability occurring in the
purely classical case. In Section IV, we also include the example of the physically relevant distribution corresponding to a bump-in-tail equilibrium. Our conclusions are given in Section
V.

\section{Quantum dispersion relations}

If $f(x,v,t)$ is the Wigner quasi-distribution and $\phi$ the scalar
potential, then the Wigner-Poisson system
\cite{Wigner}-\cite{Tatarski} reads
\begin{eqnarray}
\label{eq1}
\frac{\partial\,f}{\partial\,t} +
v\frac{\partial\,f}{\partial\,x} &=& \int\,dv'K(v' - v,x,t)f(v',x,t) \,, \\
\label{eq2}
\frac{\partial^{2}\phi}{\partial\,x^2}
&=& \frac{e}{\varepsilon_0}(\int\,dv\,f - n_{0}) \,,
\end{eqnarray}
where $K(v' - v,x,t)$ is a functional of the scalar potential,
\begin{eqnarray}
K(v' - v,x,t) &=& \frac{em}{i\hbar}\int\frac{d\lambda}{2\pi\hbar}
\exp\left(\frac{im(v' - v)\lambda}{\hbar}\right) \times \nonumber  \\
\label{eq3}
&\times& \left(\phi(x -
\frac{\lambda}{2},t) - \phi(x + \frac{\lambda}{2},t)\right) \,.
\end{eqnarray}
Here, $n_0$ is a background ionic density, $- e$ and $m$ are the
electron charge and mass, $\hbar$ is the scaled Planck constant and
$\varepsilon_0$ is the vacuum dielectric constant. We take
periodic boundary conditions in space and assume that for large $|v|$, $f$
and all its velocity derivatives tend to zero. We also
assume that the initial Wigner function is everywhere positive.
However, the time evolution determined by Eq. (\ref{eq1}) may force the
Wigner function to assume negative values.  Hence, a strict
interpretation of $f$ as a true probability distribution  is
impossible. In spite of that, the Wigner function may be used as a
useful mathematical tool to compute macroscopic
quantities such as the charge density and electric current.

The linear stability of a plasma, be it classical or quantum, is
determined by the dispersion relation, which is obtained after
Fourier transforming in space and Laplace transforming in time.
Following this procedure, we obtain \cite{Drummond} for a frequency
$\omega$ and a wave number $k$
\begin{equation}
\label{eq4}
D(k,\omega) = D_{r}(k,\omega) + i D_{i}(k,\omega) = 0 \,,
\end{equation}
where the dispersion function $D(k,\omega)$ is given by
\begin{eqnarray}
\label{eq5}
D_{r}(k,\omega) &=& 1 - \frac{\omega_{p}^2}{n_{0}k^2}\int_{P}
\frac{dv\,F(v)}{(v - \omega/k)^2 - \hbar^{2}k^{2}/4m^2} \,, \\
\label{eq6}
D_{i}(k,\omega) &=& - \,\frac{\pi\,e^2}{\hbar\varepsilon_{0}k^3}
\left(F(\frac{\omega}{k} + \frac{\hbar\,k}{2m}) - F(\frac{\omega}{k} -
\frac{\hbar\,k}{2m})\right) \,.
\end{eqnarray}
In  Eq. (\ref{eq5}), $P$ stands for the principal value symbol
and $F(v)$ denotes the (spatially homogeneous)
equilibrium Wigner function. Also, $\omega_p
= (n_{0}e^2/m\varepsilon_0)^{1/2}$ is the usual plasma frequency.

The quantum formulae reduce to the classical ones as $\hbar \rightarrow 0$.
In particular,
\begin{equation}
\label{eq7}
D_{i}(k,\omega) = - \,\frac{\pi\omega_{p}^2}{n_{0}k^2}\left(
{\frac{dF}{dv}}\right)_{v=\omega/{k}} + O(\hbar^2) \,.
\end{equation}
Moreover, no matter what the value of $\hbar$, for $|\omega|
\rightarrow \infty$ we have $D_r \rightarrow 1$ and $D_i \rightarrow
0$, as in the classical case.

The topology of the Nyquist diagram is determined by the sign of
$D_r$ at the points where $D_i = 0$. As mentioned in the
Introduction, the number of unstable modes equals the number of
times the Nyquist curve encircles the origin. Therefore, unstable modes
can only exist if $D_r <0$ for at least one of the points where $D_i =0$.
In the classical
case, the zeroes of the imaginary part of the dispersion function
are determined by the points at which the distribution function has
zero derivative. In the quantum case, according to Eq. (\ref{eq6}), the
decisive points are the real roots $v_0$ of
\begin{equation}
\label{eq8}
F(v_0 + H) = F(v_0 - H) \,.
\end{equation}
Here and in the following,
\begin{equation}
\label{eq9}
v_0 = \frac{\omega}{k} \,, \quad  H = \frac{\hbar\,k}{2m} \,.
\end{equation}
The geometrical interpretation of Eq. (\ref{eq8}) is simple: we have to
find the points $v_0$ that are equidistant to any two points at
which $F$ has the same value (see Fig. \ref{fig1}). The corresponding distance is $H$. In
a sense, Eq. (\ref{eq8}) is the finite difference version of the
classical condition $dF/dv(v = v_0) = 0$. Finally, as Nyquist's
diagram is obtained taking exclusively real frequencies, only the
real roots of  Eq. (\ref{eq8}) are relevant.

The basic tasks we have to perform are first solving
Eq. (\ref{eq8}), obtaining all real roots $v_0$ for a given $H$, and
then studying the sign of $D_r$ at each such root, taking $\omega =
k\,v_0$. Using  Eq. (\ref{eq5}), we have
\begin{equation}
D_{r}(k,\omega = kv_{0}) = 1 - \frac{\omega_{p}^2}{n_{0}k^2}
\int_{P}\,dv\,\frac{F(v)}{(v - v_{0})^2 - H^2} \,.
\end{equation}
Now, in the Cauchy principal value sense,
\begin{equation}
\int_{P}\,\frac{dv}{(v - v_{0})^2 - H^2} = 0 \,.
\end{equation}
Using this fact, we can rewrite the real part of the dispersion
function in the more convenient way
\begin{equation}
\label{eq10} D_{r}(k,\omega = kv_{0}) = 1 +
\frac{\omega_{p}^2}{n_{0}k^2} \int\,dv\,\frac{F(v_{0} + H) -
F(v)}{(v - v_{0})^2 - H^2} \,.
\end{equation}
In this form, the principal value symbol is not needed anymore,
since the integrand is regular as $v$ goes to $v_0 \pm H$. Indeed,
using the fact that $F(v_{0} + H) = F(v_{0} - H)$ from
Eq. (\ref{eq8}), we find that
\begin{equation}
\lim_{v\rightarrow{v_{0}\pm{H}}}\frac{F(v_{0} + H) - F(v)}{(v -
v_{0})^2 - H^2} =
\mp\frac{1}{2H} \frac{dF}{dv}(v_0 \pm H)
\end{equation}
is a finite quantity. A similar (but not identical) regularization procedure holds
in the classical case too \cite{Krall}.

Equations (\ref{eq8}) and (\ref{eq10}) are the fundamental
equations for Nyquist's method for one-dimensional quantum plasmas,
in which only electrostatic fields are present. So far, the
treatment has been completely general.  Let us now consider some particular
equilibria in order to analyze the consequences of Eqs. (\ref{eq8}) and
(\ref{eq10}).

\section{Equilibria with one or two maxima}

If the equilibrium Wigner function $F(v)$ has a single maximum
$v_{max}$, then the geometric meaning of $v_0$ is sufficient to
show that Eq. (\ref{eq8}) has always one, and only one, real solution
$v_0$ for any value of $H$ (see Fig. \ref{fig1}).
Depending on the shape of $F$, this
solution can differ considerably from $v_{max}$ (one has $v_0 = v_{max}$
when $F$ is symmetric with respect to $v_{max}$). However, as $H$
goes to zero, and again from geometrical  arguments, we can
convince ourselves that $v_0$ approaches $v_{max}$. Indeed
by definition $v_0$ is equidistant to the points $v'$ and
$v''$ for which $F(v') = F(v'')$. The corresponding distance from
$v_0$ to either $v'$ or $v''$ is $H$.

Furthermore, for
$(v - v_0)^2 > H^2$ we have $F(v_0 + H) > F(v)$ and for $(v -
v_0)^2 < H^2$ we have $F(v_0 + H) < F(v)$. Hence, the integrand in
 Eq. (\ref{eq10}) is always positive, implying that the real
part of the dispersion function is a positive quantity. Also, for
$|\omega| \rightarrow \infty$ we have $D_r \rightarrow 1$ and $D_i
\rightarrow 0$. Since there is only one root for
Eq. (\ref{eq8}), we deduce that the Nyquist diagram cannot encircle the
origin, and therefore no
unstable modes can exist for an equilibrium with a single
maximum. Thus, no matter how strong quantum effects are, the
conclusion is the same as for the classical case.

Let us now consider equilibria with a single minimum, $v_{min}$
(see Fig. \ref{fig2}).
This is equivalent to consider equilibrium Wigner functions with
only two maxima, as on physical grounds the equilibrium function
must decay to zero as $|v| \rightarrow \infty$.
Physically, such equilibria correspond to a situation where two
counterstreaming electron populations (with similar temperatures) co-exist.
In the classical
case, the Nyquist diagram for this situation leads to the
construction of the so-called Penrose functional
\begin{equation}
\label{eq11}
P[F] = \int\,dv\frac{F(v_{min}) - F(v)}{(v_{min} - v)^2} \,,
\end{equation}
which determines the stability properties of the system. The
inequality $P[F] < 0$ is a necessary and sufficient condition for
instability, for appropriate wave numbers.
This can be easily seen from the classical limit $H \to 0$ of Eq.
(\ref{eq10}). Classically, the points $v_0$ where $D_i = 0$ are the
maxima ($\pm v_{max}$) and the minimum ($v_{min}$) of the equilibrium distribution.
For $v_0 = \pm v_{max}$, the integrand in Eq. (\ref{eq10})
is always positive, and thus cannot lead to instability.
For $v_0 = v_{min}$, the real part of the dispersion function reduces to
$D_r = 1 +(\omega_p^2/n_0 k^2)~P[F]$. If the Penrose functional is positive,
instability is ruled out. If it is negative, one can always choose $k$ small
enough so that $D_r < 0$ and therefore some unstable modes must exist.
This completes the proof of the necessary and sufficient Penrose criterion.

The natural question now is whether there exists an analogue Penrose
functional for the quantum case.  For simplicity, in the following we restrict our discussion
to Wigner equilibria that are symmetric about $v_{min}$, the
point at which $F$ attains its minimum value. By a Galilean
transformation, this point can be taken as $v_{min} = 0$
without loss of generality.
We first notice that, in the classical case, one only has to consider
 the three velocities for which the equilibrium distribution
function has zero derivative. In the quantum case, however,
depending on the shape of the equilibrium Wigner function, there
can be more then three roots for Eq. (\ref{eq8}), with fixed $H$. For
instance, in Fig. \ref{fig2}, root $v_1$ (connecting one increasing
and one decreasing branch of
the distribution) can be obtained from the local
maximum $v_{max}$, by varying $H$ continuously from zero to a certain
value. The root $v_2$ (connecting two decreasing branches of
the distribution) is of a different nature, arising only for
sufficiently large $H$. Indeed, it is not difficult to realize
that, in the case of two maxima, there are always only three roots
for Eq. (\ref{eq8}) if $H$ is small enough, and up to five roots for larger
values of $H$. Also notice that, for symmetric equilibria,
the point $v=0$ is always a root, irrespective
of the value of $H$.
For a given $H$, possessing one, three or five real roots
depends on the details of the equilibrium.
It is not difficult to prove that, in the case of a two-humped distribution,
a sufficient (but not necessary) condition for having five roots to Eq. (\ref{eq8})
is that $F(v_{min}) = 0$. This can be shown by plotting the left- and right-hand sides
of Eq. (\ref{eq8}) as a function of $v_0$, and looking at the intersections of the
two curves. In general, we obtain that one can have five solutions when
$F(v_{min})$ is smaller than a certain threshold. Note however that five roots
only appear for sufficiently large values of $H$; for small enough $H$, there are
always only three roots.
In Section IV, we shall examine
a bi-Lorentzian distribution possessing at most three solutions. In
addition, we shall discuss another two equilibria, which possess
five solutions for sufficiently large $H$.

Let us now consider the question of the existence of a quantum
Penrose functional. We need to examine the sign of $D_r$
at the different solutions of Eq. (\ref{eq8}).
The root $v_0 = v_{min}=0$ always exists and can yield either a positive or a negative
value for the integral in Eq. (\ref{eq10}), depending on the shape of
the equilibrium and the value of $H$.  One can actually prove that  the
integral can be negative only if $H < v^\star$, where $v^\star$ is the
positive solution of the equation $F(0)=F(v^\star)$.

We now analyze the other roots of Eq. (\ref{eq8}).
Let $v_1$ be the root
obtained from the maximum of $F$ at the right of $v = 0$ by varying
continuously $H$ from zero to some particular value (see Fig. \ref{fig2}).
Referring to Fig. \ref{fig2} and to  Eq. (\ref{eq10})
(with $v_0 = v_1$), we conclude that
the integrand in $D_r$ is negative for $- v_1 - H < v < - v_1 + H$.
Thus, in principle, the real part of the dispersion function can be
negative. However, one could imagine that the negative contribution for $- v_1
- H < v < - v_1 + H$ is compensated by a positive contribution
corresponding to $v_1 - H < v < v_1 + H$. Let us examine this
possibility. Using the fact that $F$ is even, we obtain
\begin{eqnarray} \label{eq12} \int_{-v_1-H}^{-v_1+H}\,&dv&
\frac{F(v_{1}+H)-F(v)}{(v-v_{1})^{2}-
H^2}+\int_{v_1-H}^{v_1+H}\,dv\,\frac{F(v_{1}+H)-F(v)}{(v-v_{1})^{2}-H^2}=
\nonumber \\
&& \\
&=&2\int_{v_1-H}^{v_1+H}\,dv\,\frac{(F(v_{1}+H)-F(v))(v^{2}+v_{1}^{2}-
H^2)}{(v^{2}-(H+v_1)^2)(v^2-(H-v_1)^2)}\,. \nonumber
\end{eqnarray}
For $v_1 - H < v < v_1 + H$, we have $F(v) > F(v_1 +
H)$, $v^2 > (H - v_1)^2$ and $v^2 < (H + v_1)^2$. Hence, the
integrand in Eq. (\ref{eq12}), which can give the only negative
contribution for $D_r$, is negative provided
\begin{equation}
\label{eqq12}
v^2 < H^2 - v_{1}^2 \,,
\end{equation}
which is impossible in the prescribed range of velocities, as $v_1 > H$
by construction.
Therefore, we always have  $D_{r}(k,\omega = kv_{1}) > 0$, where
$v_1$ is the (semi-classical) root for Eq. (\ref{eq8})
obtained from the positive maximum of $F$, and the same argument
holds for the symmetric root $-v_1$. This is analogous to the classical
result shown above, according to which $D_r$ is positive at the two maxima
of $F(v)$. Indeed, the roots $\pm v_1$ coincide with $\pm v_{max}$
when $H \to 0$.

However, this is not the end of the story for the quantum case.
Indeed, for sufficiently large values of $H$, it is possible to
access the roots $\pm v_2$ (connecting two decreasing branches of
the distribution) shown on Fig. \ref{fig2}. [This is not in
contradiction with the above statement that some equilibria only
display three solutions to Eq. (\ref{eq8}). Solutions of the type
$v_2$ always exist, although they may correspond to {\it different
values of} $H$ than $v_1$, so that for a fixed $H$ there are
indeed only three roots]. For the roots $\pm v_2$, which are of a
strictly quantum nature, we cannot anymore obtain, a priori, $D_r
> 0$. For instance, for the particular choice of $v_2$ shown on
Fig. \ref{fig2}, the region $ - v_2 - H < v < - v_2 + H$
contributes a negative value to $D_{r}(k,\omega = kv_2)$. The same
is true for the root $- v_2$. This is because, over most of the
region $ - v_2 - H < v < - v_2 + H$ one has $F(v_2 +H) <F(v)$ and
$(v-v_2)^2 > H^2$, implying that the integral in Eq. (\ref{eq10})
is negative. Another choice of $v_2$ may have yielded the opposite
result, so that the sign of $D_{r}(k,\omega = kv_2)$ cannot be
determined a priori. As the parameter $H$ depends on the wave
number, it is always possible to choose $k$ so as to access a root
of the type $\pm v_2$, for which the sign of $D_r$ is
undetermined. The conclusion is that there is {\it no} quantum
Penrose functional, since the topology of the Nyquist diagram can
be changed, in an essential way, by the value of $D_r$ at the
quantum roots for Eq. (\ref{eq8}). Each specific equilibrium must
be studied in detail.  In the following Section, we shall illustrate the
previous theory using some concrete examples.

\section{Examples of two-stream and bump-in-tail equilibria}

Let us consider a two-humped equilibrium given by
\begin{equation}
\label{eq14}
F(v) = \frac{n_{0}\Delta}{2\pi}\left(\frac{1}{(v - a)^2 + \Delta^2} +
\frac{1}{(v + a)^2 + \Delta^2}\right) \,,
\end{equation}
where $\Delta$ is a measure of the dispersion of the distribution
and $a$ is a parameter associated to the distance between the two
possible maxima. If $a^2 < \Delta^{2}/3$ this bi-Lorentzian
distribution degenerates into a one-humped equilibrium, which is
consequently stable against linear perturbations, both in classical
and quantum cases. The major advantage of dealing with Eq. (\ref{eq14})
is that it is amenable to exact calculations, thus providing an
appropriate example of the use of the Nyquist method for quantum
plasmas. Moreover, it models the physically relevant situation of two
counterstreaming electron populations that co-exist within the
same plasma.

Inserting Eq. (\ref{eq14}) into Eq. (\ref{eq8}), we obtain the following solutions,
\begin{eqnarray}
\label{eq15}
v_{0}^0 &=& 0 \,, \\
\label{eq16}
v_{0}^1 &=& \pm\left(H^2 - a^2 - \Delta^2 + 2\sqrt{a^2 +
\Delta^2}\sqrt{a^2 - H^2}\right)^{1/2} \,,\\
\label{eq17}
v_{0}^2 &=& \pm\left(H^2 - a^2 - \Delta^2 - 2\sqrt{a^2 +
\Delta^2}\sqrt{a^2 - H^2}\right)^{1/2} \,.
\end{eqnarray}
It is easy to check that the roots given in Eq. (\ref{eq17}) are always complex,
whatever the values of $H$, $a$ and $\Delta$. However, the roots
(\ref{eq16}) can be real, provided
\begin{equation}
\label{eq18}
a^2 > \frac{1}{3}(H^2 - \Delta^2) + \frac{2}{3}(\Delta^4 + H^{2}\Delta^2 +
H^4)^{1/2} \,.
\end{equation}
Thus, there can be one or three relevant roots, according to
condition (\ref{eq18}). This inequality, when satisfied, can also
be seen to imply $a^2 > \Delta^2/3$, which is the same as the
condition for the existence of two maxima. Hence, there can be three
real roots if and only if $F$ is two-humped, which is not
surprising in view of the arguments given in the preceding Section.

An equivalent and illuminating way to rewrite Eq. (\ref{eq18}) is
\begin{equation}
\label{eqqqq18}
H^2 < v_{max}^2 \,,
\end{equation}
where $v_{max}$ denotes the (positive) point where $F$ is maximum,
\begin{equation}
\label{eqq18}
v_{max} = (a^2 + \Delta^2)^{1/4}\left(2a - \sqrt{a^2 + \Delta^2}\right)^{1/2} \,.
\end{equation}
Hence, $2H$ cannot exceed the distance between the two maxima of
$F$. Notice that the right hand side of Eq. (\ref{eqq18}) can be
real only if $a^2 > \Delta^2/3$, that is, if there are two
maxima, which is again a natural result.  For very large quantum
effects, only the root $v_{0}^0 = 0$ survives.

As there is no quantum Penrose functional, it is necessary to
calculate $D_r$ at all possible roots (\ref{eq15})-(\ref{eq16}). We
obtain
\begin{equation}
\label{eq19} D_{r}(k, \omega = kv_{0}^0 = 0) = 1 +
\frac{\omega_{p}^2}{k^2}\frac{ (\Delta^2 + H^2 - a^2)}{(H^2 -
a^2)^2 + 2\Delta^2(H^2 + a^2) + \Delta^4} \,,
\end{equation}
which can be negative if and only if
\begin{equation}
\label{eq20}
a^2 > \Delta^2 + H^2 \,.
\end{equation}
In addition,
\begin{equation}
\label{eq21}
D_{r}(k, \omega = \pm kv_{0}^1) = 1 +
16\,\frac{\omega_{p}^{2}\sqrt{a^2 + \Delta^2}(a^2 -
H^2)a^{2}\delta}{k^{2}U^8} \,,
\end{equation}
where $\delta$ and $U^8$ are the positive-definite quantities
\begin{eqnarray}
\delta &=& \sqrt{a^2 + \Delta^2} - \sqrt{a^2 - H^2} \,, \\
U^8 &=& \left(((v_0 - H)^2 - a^2)^2 + 2\Delta^2((v_0 - H)^2 + a^2) +
\Delta^4\right) \times \nonumber \\
\label{eq22}
&\times& \left(((v_0 + H)^2 - a^2)^2 + 2\Delta^2((v_0 + H)^2 + a^2) +
\Delta^4\right) \,.
\end{eqnarray}
We can show that Eq. (\ref{eq18}) implies $a^2 > H^2$, so that $D_r$ as given by
Eq. (\ref{eq21}) is indeed always positive.

In view of Eqs. (\ref{eq19}) and (\ref{eq21}), we see that Eq. (\ref{eq20})
is a necessary and sufficient condition for linear instability.
That this condition is sufficient can be easily proven: suppose that we have found
a wave number $k_0$ satisfying Eq. (\ref{eq20}); then any $k < k_0$ will also
satisfy it; by taking $k$ small enough, we can make the second addendum of Eq. (\ref{eq19})
(which is negative) arbitrarily large in absolute value and therefore
obtain $D_r <0$. Note, however, that putting an equality sign in Eq. (\ref{eq20})
and solving for $k$ does {\it not} provide the transition wave number between
stable and unstable behavior. In order to obtain it, one has to set Eq. (\ref{eq19})
to zero and solve for $k$.

Equation (\ref{eq20}) means that the plasma can become unstable for sufficiently
large $a$ (the two maxima are sufficiently far apart), small
$\Delta$ (small dispersion) or small $H$ (small quantum effects).
We also notice that, as $H$ depends on the wave number,
quantum effects can suppress the instability for small wavelengths.
The instability condition Eq. (\ref{eq20}) confirms the numerical
results by Suh {\it et al.} \cite{Suh}. Here, however, we have derived an
{\it exact} analytical criterion for quantum linear stability of a
two-stream equilibrium.

On Fig. \ref{fig3}, we have plotted the Nyquist diagrams for the two-stream equilibrium
of Eq. (\ref{eq14}) with $a=3$, $\Delta=1$, $k=0.2$ and four different
values of $\hbar$ (units for which $e=\varepsilon_0 =m=n_0 = 1$ are used).
We observe that stabilization of the $k=0.2$ mode
occurs somewhere between $\hbar=25$ and $\hbar = 27$.
This is in agreement with the previous formulae: indeed, with this set of
parameters, it is found that $D_r(k,\omega=kv_0^0)$ changes sign for $\hbar \simeq 25.5$.
Furthermore, Fig. \ref{fig3} also shows a change in the topology of the Nyquist diagram.
For figures (a)$-$(c), the diagram intersects the horizontal axis in three points
[excluding the point (1,0) that corresponds to $\omega =\pm \infty$]; note that two such
points coincide, because of the symmetry of the distribution.
For figure (d), only one intersection survives.
This change in topology corresponds to having one or three solutions to Eq. (\ref{eq8}),
which is determined by Eq. (\ref{eq18}). The transition is found to occur for
$\hbar \simeq 30$, which is in agreement with the diagrams of Fig. \ref{fig3}.

Finally, we point out that large quantum effects are not necessarily
stabilizing. For the two-stream equilibrium
of Eq. (\ref{eq14}), with $a=3$ and $\Delta=1$, the wave number $k=0.287$ is
classically stable. However, increasing $\hbar$, one finds that $D_r(k,\omega=kv_0^0)$
becomes negative on an interval approximately given by $7 < \hbar < 11.8$.
This can be easily verified by plotting Eq. (\ref{eq19}) as a function of $\hbar$
or by direct substitution of the above values. However, this destabilizing
effect occurs for rather
limited ranges of $\hbar$ and $k$. For example, wave numbers $k<0.28$
are classically unstable and are stabilized for large enough $\hbar$ (as in our previous
example with $k=0.2$); on the other hand, wave numbers $k>0.29$ are classically stable
and remain stable for any value of $\hbar$. Only wave numbers very close to the value
$k=0.287$ display the unusual behavior described above.
For this reason, we can still conclude that
the most likely outcome of quantum effects is stabilization.

We now show that we can explicitly write a distribution function for which there can
exist five real roots for Eq. (\ref{eq8}).  Consider the two-humped
equilibrium
\begin{equation}
\label{eq23}
F(v) = \frac{2n_0}{\sqrt{\pi}a^3}v^{2}\exp(- v^{2}/a^2) \,,
\end{equation}
where $a$ is a parameter related to the equilibrium temperature.
As $F(0) = 0$, we should expect that Eq. (\ref{eq8}) possesses five
real roots for some (large enough) $H$. We now give an explicit proof
of this fact for the above equilibrium.
The solutions to Eq. (\ref{eq8}) are obtained in this case from
the equation
\begin{equation}
\tanh\left(\frac{2Hv}{a^2}\right) = G(v;H) \,,
\label{gaussian}
\end{equation}
where we have defined
\begin{equation}
G(v;H) = \frac{2Hv}{v^2 + H^2} \,.
\end{equation}
Apart from the obvious root $v = 0$, we can have two or four
additional real roots. By plotting the left-
and right-hand side of Eq. (\ref{gaussian}) as a function of $v$
(see Fig. \ref{fig4}),
we can show that there will be a total of five real roots if and
only if
\begin{equation}
\label{eq24}
\frac{dG}{dv}(v=0) < \frac{d}{dv}\tanh\left(\frac{2Hv}{a^2}\right)(v=0) \,.
\end{equation}
This implies that there can be five real roots provided $H > a$,
that is, for sufficiently large quantum effects. Otherwise, only
three (semi-classical) solutions are possible.

In the remaining part of this Section, we address the question of
the quantum linear stability of an equilibrium characterized by a
large central distribution of electrons with in addition a small
bump in the tail. This is a standard problem in plasma physics,
with the small perturbation to the central distribution
representing a beam injected in the plasma. Here, we consider the
quantum aspects of the problem, by using Nyquist diagrams. The
so-called bump-in-tail equilibrium has a single minimum, but, as
there is no quantum Penrose functional, we are lead to compute the
real part of the dispersion function at all critical points
(zeroes of the imaginary part of the dispersion function).
Nevertheless, the Nyquist technique is less expensive than, for
instance, direct calculation of the dispersion relation, since it
requires the value of $D_r$ at a few points only.

To model the bump-in-tail equilibrium, we use the following
distribution (see Fig. \ref{fig5})
\begin{equation}
F = \frac{2n_0}{3\pi\,a}\frac{[1-\sqrt{2}(v/a)]^2}{[1 +
(v/a)^2]^2} \,,
\label{bump}
\end{equation}
where $a > 0$ is a reference velocity that can be scaled to unity
without loss of generality. Henceforth, we set $a =1$. The
distribution of Eq. (\ref{bump}) is a particular case of a one-parameter
family of bump-in-tail equilibria whose classical linear stability
properties have been recently studied via Nyquist diagrams
\cite{Diego}. In the quantum case, there is no Penrose functional
and the analysis is more involved.

Inserting Eq. (\ref{bump}) into the determining equation Eq.
(\ref{eq8}), we obtain the following equation for $v_0$
\begin{equation}
\frac{1-\sqrt{2}(v_0+H)}{1 +(v_0+H)^2} = \pm
\frac{1-\sqrt{2}(v_0-H)}{1 +(v_0-H)^2} \,.
\label{xxx}
\end{equation}
The plus sign yields the second-degree equation :
$v_0^2 -\sqrt{2} v_0 -H^2 =0$, with solutions
\begin{equation}
v_{\pm}(H) = {1 \over 2} \left( \sqrt{2}\pm \sqrt{6+4 H^2} \right)\,.
\label{quadr}
\end{equation}
Note that, in the limit $H \to 0$, these solutions correspond to the two
maxima of the equilibrium distribution.
Taking the minus sign in Eq. (\ref{xxx}) yields the third-degree equation
\begin{equation}
\sqrt{2} v_0^3 -v_0^2 + \sqrt{2}(1-H^2) v_0 - (1+H^2) =0 ~.
\label{cubic}
\end{equation}
It is easy to prove that Eq. (\ref{cubic}) has one real solution for
$H<3$ and three real solutions for $H \ge 3$. Furthermore, the largest of such solutions
is always positive, and coincides with the minimum of the equilibrium distribution
when $H=0$ : we shall call this solution $v_{m}(H)$.
The other two solutions (which are real only when $H \ge 3$) have no classical
counterpart, and will be called $v_{q1}(H)$ and $v_{q2}(H)$.
A graph of all the roots of Eq. (\ref{xxx}) as a function
of $H$ is provided on Fig. \ref{fig6}.
Again, the existence of five real roots for some values of $H$ is a
consequence of the fact that $F(v_{min}) = 0$.

As further calculations are rather cumbersome,
we only report here the most relevant results
(mostly obtained using the mathematical package {\it MAPLE}).
For $H \ge 3$, we have obtained numerically,
using Eq. (\ref{eq10}), that $D_{r}(k,\omega = kv_{q1}) > 0$ and
$D_{r}(k,\omega = kv_{q2}) > 0$. At least for this particular
example, this can be shown to imply that the
purely quantum solutions are irrelevant to the linear stability
properties of the equilibrium. Moreover, we have found numerically that
$D_{r}[k,\omega = kv_-(H)] > 0$. After an involved analysis to
determine the ordering of all the solutions for Eq. (\ref{eq8}),
the conclusion is that the unstable modes satisfy
\begin{equation}
\label{x2} D_{r}[k,\omega = kv_{m}(H)] < 0 \,, \quad
D_{r}[k,\omega = kv_{+}(H)] > 0 \,.
\end{equation}
This holds whatever the value of $H$. Remembering the dependence
of $H$ on the wave-number and taking into account the explicit
forms of $v_{m}(H)$ and $v_{+}(H)$, it appears that the
pair of conditions (\ref{x2}) are very complicated expressions of
$k$ and $\hbar$. \ However, using appropriated units in which $n_0
= \omega_p = m = 1$, we were able to solve Eq. (\ref{x2}) numerically
for a few values of $\hbar$ (measured in units of
$ma^2/\omega_p$). For $\hbar = 0$, we found that the unstable
modes satisfy $0.18 < k < 0.82$, where $k$ is measured in units of
$\omega_{p}/a$. This is the classical condition
for linear instability. For $\hbar = 10$, the instability range is given
by $0.20 < k < 0.36$. We see that the total band of instability
becomes smaller for a non-zero Planck's constant. Further
increasing $\hbar$, taking $\hbar = 100$, we found that the
unstable linear waves must satisfy $0.15 < k < 0.17$. For even
larger quantum effects, there is virtually a suppression of all
unstable modes. This is again in agreement with the numerical results of Suh {\it
et al.} \cite{Suh}, where large quantum effects were shown to
stabilize all classically unstable modes for a two-stream equilibrium.

\section{Conclusion}

In this paper, we have discussed the Nyquist method for the study of
the linear stability of spatially homogeneous
quantum plasmas described by the Wigner-Poisson system.
For classical Vlasov-Poisson plasmas, this method provides a simple
way to analyze the stability properties. Furthermore, for the special case
of two-stream equilibria, one can construct a simple functional (known as
Penrose functional), whose sign determines whether unstable modes exist.

The main conclusion of the present work is that the stability analysis of
quantum plasmas is generally subtler than in the classical case.
In particular, we have shown that no simple analogue of the Penrose
functional can be constructed in order to
determine the stability properties of a
two-humped equilibrium. Hence, a detailed analysis is necessary for each
particular case, with generic and universal conclusions being more
difficult to obtain.
However, we were able to prove that
one-humped equilibria (i.e., with a distribution that is a monotonically
decreasing function of the energy) are always stable: this is the same
result as for classical plasmas.

The main mathematical reason for the subtler behavior
of quantum plasmas is that the wave number
now enters the real part of the dispersion function through the
parameter $H=\hbar k/2m$. This can change the topology of the Nyquist
diagram, not only by varying $\hbar$, but also by varying the wave number at
fixed $\hbar$. Physically, this means that new unstable modes can arise
by resonant interaction between the quantum velocity $H$ and some other
typical plasma velocity. Indeed, such purely quantum unstable modes
have been observed \cite{Haas} for the special case of two-stream
equilibrium
\begin{equation}
F(v) = {n_0 \over 2} \delta(v-a) + {n_0 \over 2} \delta(v+a) ~,
\end{equation}
where $\delta$ is the Dirac delta function and $\pm a$ the velocities of each stream.
This equilibrium can be amenable to exact calculation \cite{Haas}.

Even when general or exact results cannot be obtained,
the Nyquist technique can be successfully used for the
study of particular equilibria, as was shown in Section IV. The
bi-Lorentzian equilibrium treated in that Section has shown that large quantum effects
generally contribute to stabilize perturbations \cite{Suh}.
This is not always the case, however, and we have produced an explicit
example of a wave number that is classically stable and becomes unstable for finite $\hbar$.
Moreover, the Nyquist
method has enabled us to derive an exact stability criterion for such a
bi-Lorentzian equilibrium. The
Nyquist technique was also applied to a classically unstable bump-in-tail equilibrium.
Again, large quantum effects were shown to reduce the range of unstable
wave numbers.

\vskip 1cm
\noindent{\bf Acknowledgments}\\
We are grateful to P.~Bertrand for valuable comments and
suggestions. One of us (F. H.) thanks the Laboratoire de Physique
des Milieux Ionis\'es for hospitality while part of this work
was carried out and the Brazilian agency Conselho Nacional de
Desenvolvimento Cient\'{\i}fico e Tecn\'ologico (CNPq) for
financial support.

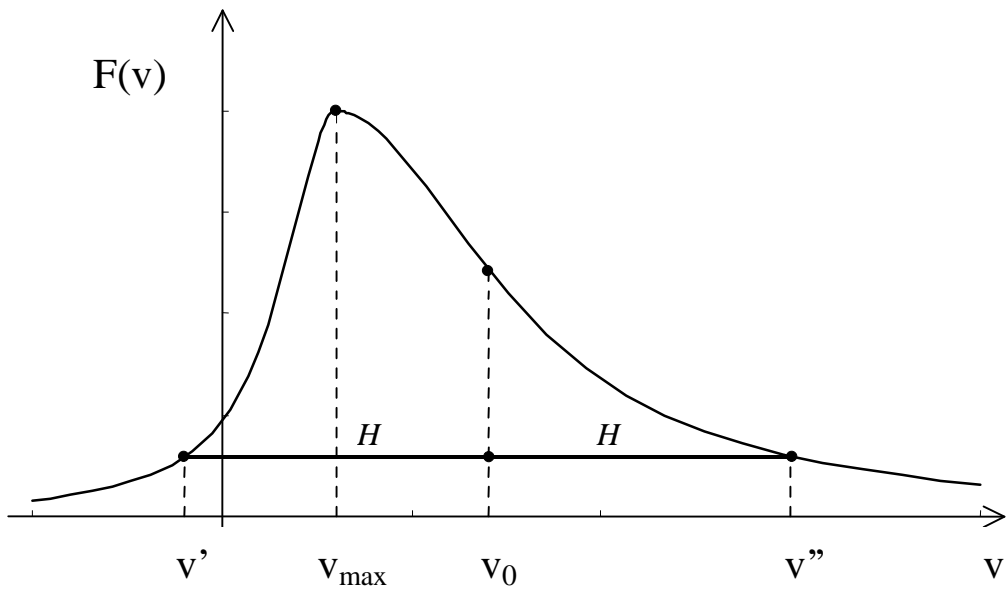
\begin{figure}
\caption{
Graphical representation of the geometric meaning of $v_0$
[solution of Eq. (\ref{eq8})] for a one-humped distribution
function.
The distance between $v_0$ and both $v'$ and $v''$ is
equal to $H$. The Wigner function is
represented on the vertical axis and the velocities on the horizontal axis.
}
\label{fig1}
\end{figure}
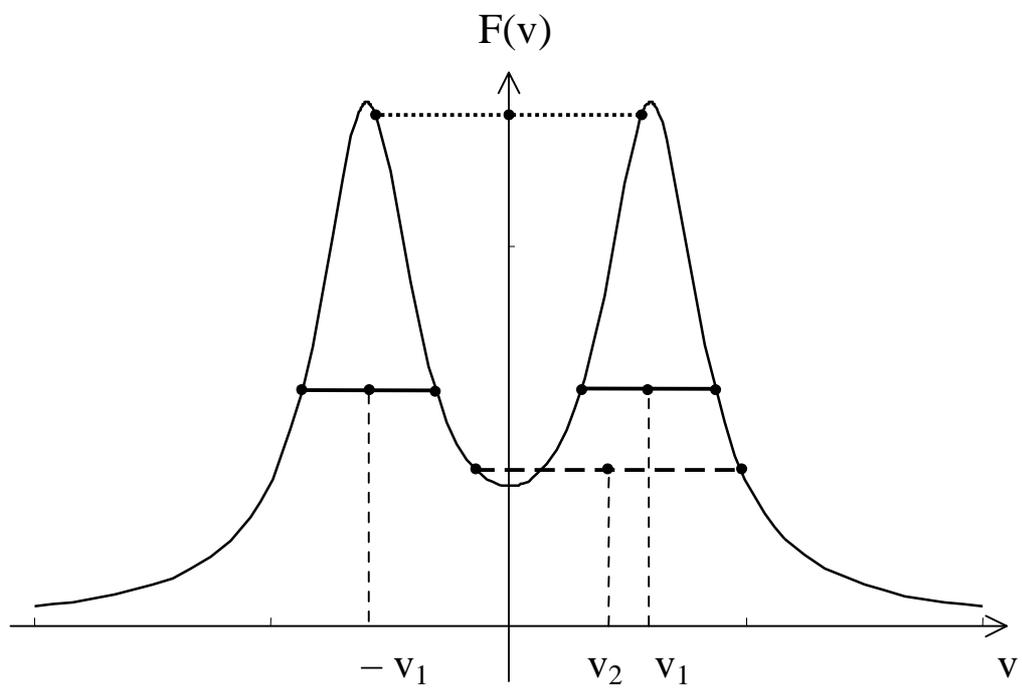
\begin{figure}
\caption{
Semi-classical ($v_0 = \pm v_1$, solid horizontal lines)
and purely quantum ($v_0 = \pm v_2$, dashed horizontal line) solutions of
Eq. (\ref{eq8}) for a symmetrical two-stream equilibrium. Also note
that $v_0 = 0$ (dotted line) is always a solution.
Units are conveniently rescaled.
}
\label{fig2}
\end{figure}
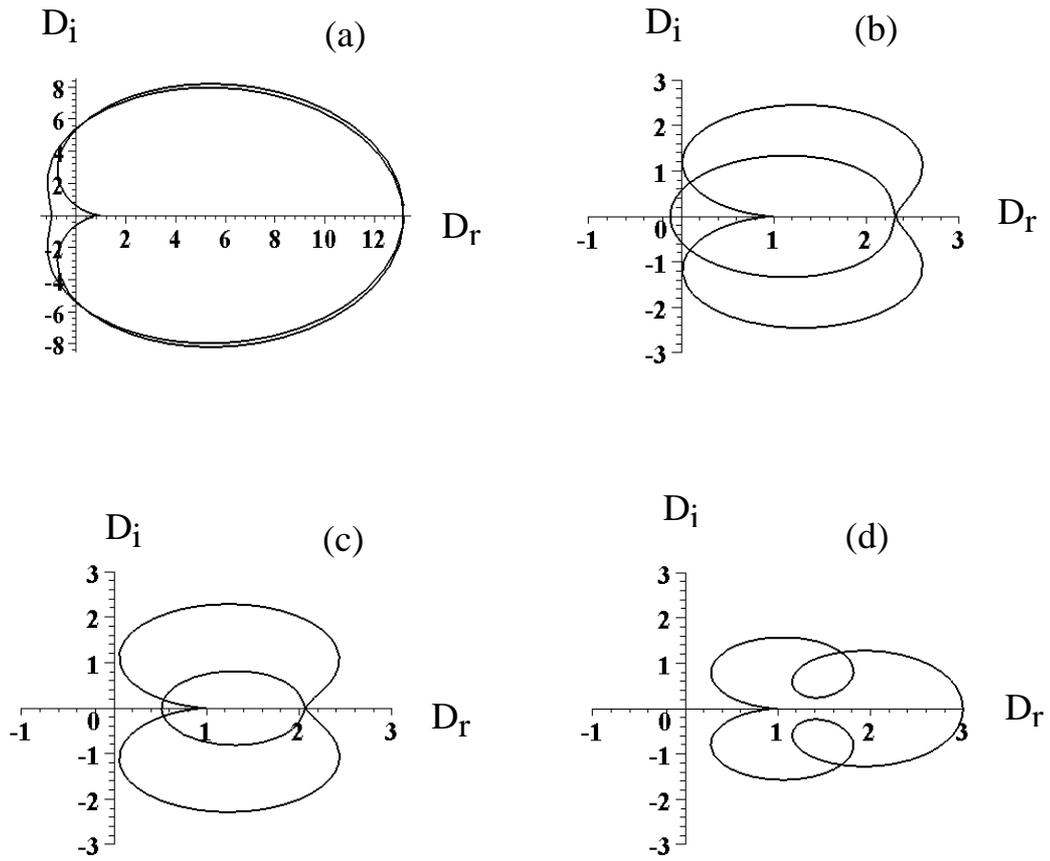
\begin{figure}
\caption{ Nyquist diagrams for the two-stream equilibrium of Eq.
(\ref{eq14}) with $a=3$, $\Delta=1$, $k=0.2$ and $\hbar = 0.001$
(a), 25 (b), 27 (c) and 40 (d) (units for which $e=\varepsilon_0 =m=n_0 = 1$ are used).
Diagrams (c) and (d) indicate that quantum effects have suppressed
the instability.
} \label{fig3}
\end{figure}
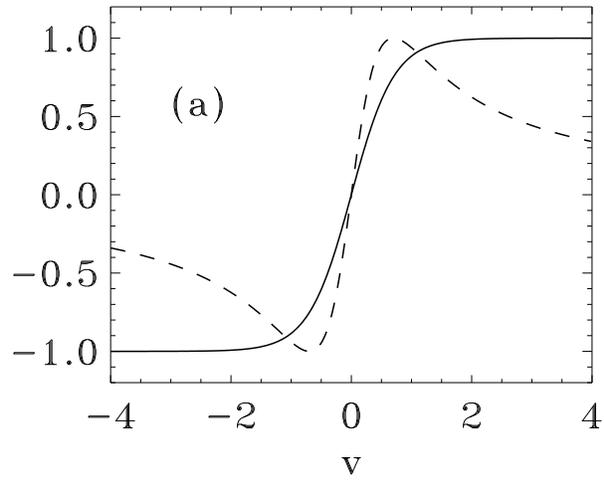
\begin{figure}
\caption{ Plot of the left-hand side (solid line) and right-hand
side (dashed line) of Eq.
(\ref{gaussian}) as a function of $v$, for $a=1$ and $H = 0.7$ (a)
and $H = 1.2$ (b).
The inset is a zoom of the region $0.9 <v<1.4$ for case (b),
showing in detail
the extra solutions arising for $H > a$.  } \label{fig4}
\end{figure}
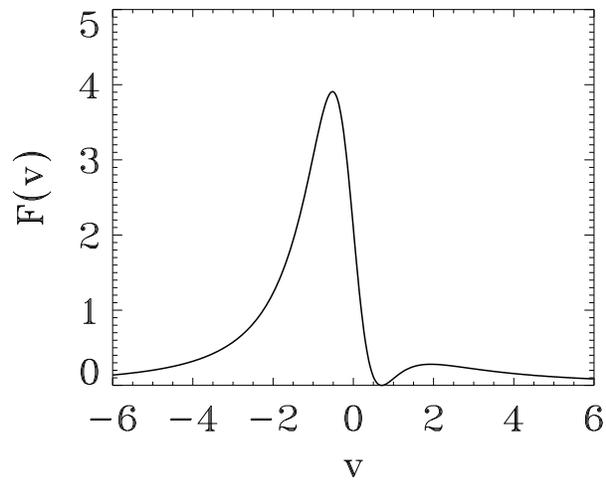
\begin{figure}
\caption{ Velocity distribution corresponding to the bump-in-tail
equilibrium of Eq. (\ref{bump}) for $a=n_0=1$. } \label{fig5}
\end{figure}
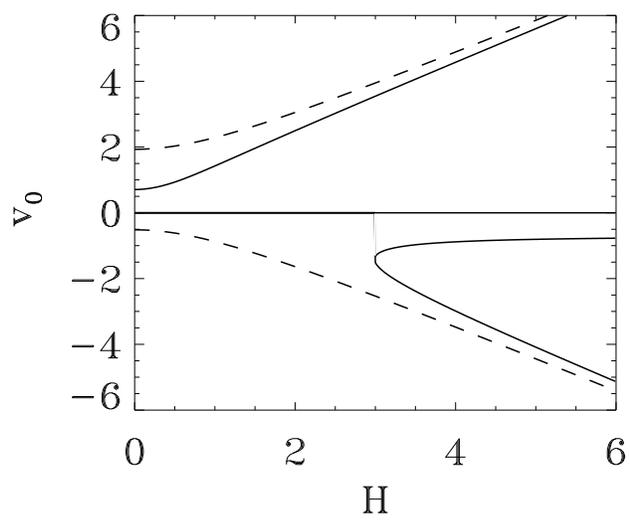
\begin{figure}
\caption{
Plot of the roots $v_0$ of Eq. (\ref{xxx}) as a function
of $H$. The dashed lines represent the roots $v_\pm$ in Eq. (\ref{quadr});
the solid lines represent solutions of the cubic equation (\ref{cubic}).
} \label{fig6}
\end{figure}

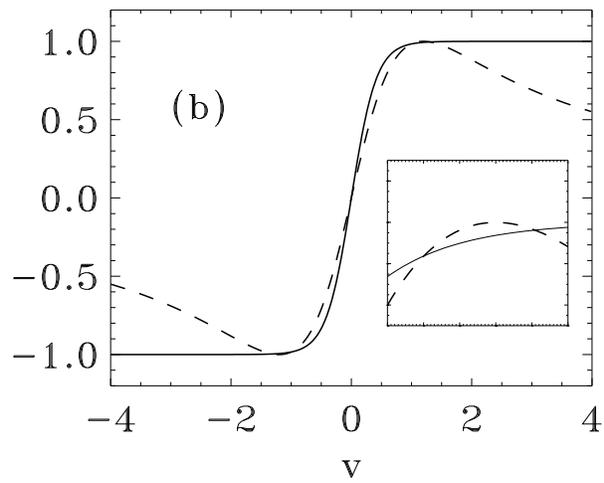
\end{document}